\documentclass[12pt,reqno]{amsart}
\usepackage{cite}
\newcommand{\M}{{\bf M}}

\newcommand{\be}{\begin{equation}}
\newcommand{\ee}{\end{equation}}
\newcommand{\bi}{\begin{itemize}}
\newcommand{\ei}{\end{itemize}}
\newcommand{\bea}{\begin{eqnarray}}
\newcommand{\eea}{\end{eqnarray}}
\newcommand{\ba}{\begin{array}}
\newcommand{\ea}{\end{array}}
\newcommand{\bt}{\begin{tabular}}
\newcommand{\et}{\end{tabular}}
\newcommand{\bc}{\begin{center}}
\newcommand{\ec}{\end{center}}

\newcommand{\h}{\hat h}
\newcommand{\K}{\hat K}
\newcommand{\J}{\hat J}
\def\vev#1{\langle #1\rangle}
\begin{document}
\begin{titlepage}
\begin{center}

\hfill hep-th/0203138\\

\vskip 3cm
{\large \bf Aspects of Spontaneous $N=2\to N=1$\\ Breaking
in Supergravity}\footnote{Work
supported by:
DFG -- The German Science Foundation,
GIF -- the German--Israeli
Foundation for Scientific Research,
European RTN Program HPRN-CT-2000-00148 and the
DAAD -- the German Academic Exchange Service.}

\vskip 1cm

{\bf Jan Louis\footnote{email: 
{\tt j.louis@physik.uni-halle.de}} }  \\

\vskip 0.6cm
{\em Fachbereich Physik, Martin-Luther-Universit\"at Halle-Wittenberg,\\
Friedemann-Bach-Platz 6, D-06099 Halle, Germany}

\end{center}

\vskip 1.5cm

\begin{center} {\bf ABSTRACT } \end{center}
We discuss 
some issues related to spontaneous $N=2\to N=1$ supersymmetry breaking.
In particular, we state a set of geometrical conditions
which are necessary that such a breaking occurs.
Furthermore,
we discuss the low energy $N=1$ effective Lagrangian and show that
it satisfies non-trivial consistency conditions which
can also be viewed as conditions on the geometry
of the scalar manifold.

\vskip 1.5cm
\begin{center} 
{\it Talk given at the\\
 ``Workshop on special geometric structures
in string theory'',\\
 Bonn, September 8-11, 2001}\end{center}

\vfill

March  2002
\end{titlepage}

\section{Introduction}

The possibility of spontaneous $N=2\to N=1$ supersymmetry breaking 
in four space-time dimensions ($D=4$)
was first  discussed in string theory \cite{HP} and 
then further developed in global supersymmetry and supergravity
(see, for example, refs.\ 
\cite{BG,APT,FGP,FGPT,Bac,JM,KK,TV,PM1,CKLT}).\footnote{%
The truncation of $N=2\to N=1$ theories has been worked out in
\cite{ADF} and aspects about partial supersymmetry breaking for
$N>2$ has been discussed in \cite{ADFL}. }
In this talk we focus on the situation in supergravity 
where so far only a few models with spontaneous $N=2\to N=1$ 
breaking are known \cite{FGP,FGPT}.
Thus it is of interest to uncover the general story of
spontaneous $N=2\to N=1$ breaking or in other words 
ask the questions

\bi
\item[(i)]
Under what (geometrical) conditions does the $N=2$ 
theory have ground states that are $N=1$ supersymmetric.

\item[(ii)]
What is the low energy effective $N=1$ action which describes the interactions
below the scale 
of supersymmetry breaking.
\ei

As we will see the first question (i)
can be rephrased  as a geometrical conditions
on the scalar manifold which is spanned by the 
scalar fields in vector- and hypermultiplets.
The second question (ii)  imposes a set of consistency conditions
on the couplings of both the original $N=2$ theory and the
low energy effective $N=1$ theory. Again these conditions can be stated 
as geometrical properties of the scalar manifold.
In this talk we address both issues and give some partial results
but leave a 
more detailed and complete analysis to a separate publication
\cite{GL}.

\section{The starting point: gauged $N=2$ supergravity}
%
Let us first briefly recall the starting point of our analysis.
In $D=4$ the spectrum of a generic $N=2$ theory consists of a 
gravitational multiplet,
$n_v$ vector multiplets and $n_h$ hypermultiplets \cite{N2}.
The gravitational multiplet 
$(g_{\mu\nu},\Psi_{\mu A}, A_\mu^0)$ 
features the space-time metric
$ g_{\mu\nu}, \mu,\nu =0,\ldots,3$, 
two gravitini $\Psi_{\mu A}, A=1,2$ and the
graviphoton $A_\mu^0$.
A vector multiplet $(A_\mu,\lambda^{A}, z)$
contains a vector  
$A_\mu$, two gaugini $\lambda^{A}$  and a complex scalar $z$.
Finally, a hypermultiplet $(\zeta_{\alpha}, q^u)$
contains two hyperini $\zeta_{\alpha}$
and 4 real scalars $q^u$. For an arbitrary number of vector-
and hypermultiplets there is a total of 
$2n_v +4n_h$ real scalar fields in the spectrum with
$\sigma$-model type interactions of the form
\be\label{sigmaint}
{L}\ =\ -g_{i\bar j}(z,\bar z)\, D_\mu z^i D_\mu\bar z^{\bar j}
- h_{uv}(q)\, D_\mu q^u D_\mu q^v +\ldots\ ,
\end{equation}
where the range of the indices is
$i,\bar j= 1,\ldots,n_v$ and $u,v=1,\ldots,4n_h$.
The scalars  $(z^i,q^u)$ can be viewed as coordinates of the manifold
\be
{\M} = {\M}_v\times {\M}_h \ ,
\end{equation}
where $g_{i\bar j}$ is the metric of the $2n_v$-dimensional space
${\M}_v$ while $h_{uv}$ is the metric on the $4n_h$-dimensional space 
 ${\M}_h$.
$N=2$ supersymmetry imposes that ${\M}_v$ is a 
special K\"ahler manifold \cite{DP,CRTP} while 
${\M}_h$ has to be a quaternionic-K\"ahler manifold \cite{BW}.

Both sets of scalar fields can be charged under the isometries of
${\M}$
\be
\delta q^u \ = \ \Lambda^I k_I^u(q)  \ ,\qquad 
\delta z^i \ = \ \Theta^I k_I^i(z)  \ ,\qquad I=0,\ldots,n_v\ ,
\end{equation}
where $ k_I^u(q), k_I^i(z)$ are Killing vectors of ${\M}_h$ and
${\M}_v$, respectively, and $\Lambda^I,\Theta^I$ are the 
respective gauge parameters. This in turn fixes the covariant derivatives
to be
\be
D_\mu q^u = \partial_\mu q^u +k_I^u A_\mu^I\ ,\qquad
D_\mu z^i = \partial_\mu z^i +k_I^u A_\mu^I\ .
\end{equation}
In the following we are mainly interested in the case where $k_I^i=0$ 
and for simplicity we focus on this situation henceforth.

On a quaternionic manifold the Killing equation 
$\nabla_u k_v +\nabla_v k_u =0$ determines the Killing vectors
in terms of a triplet of Killing prepotentials $P^x(q), x=1,2,3$ \cite{DFF}
\be\label{Pdef}
k^u_I\,K_{uv}^x \ = \  D_v P_{I}^x\ ,
\end{equation}
where $K_{uv}^x$ is the triplet of covariantly constant hyper-K\"ahler 
two-forms which exist on a quaternionic-K\"ahler manifold. 
They are related
to the triplet of complex structures $J^x$ 
(which satisfy the quaternionic algebra 
$J^x J^y = -\delta^{xy}{\bf 1} + \epsilon^{xyz} J^z$)
via
\be
K^x_{uv} = h_{uw} (J^x)^w_v \ .
\end{equation}
%
$D_v$ in (\ref{Pdef}) 
is a covariant derivative with respect to the $Sp(1)$ connection
of the holonomy group $Sp(1)\times Sp(2n_h)$.

$N=2$ supersymmetry determines the Lagrangian and thus the interaction
of the various multiplets \cite{DP,BW,DFF,N2,CRTP}.
Here we do not recall all the couplings but only focus
on those terms which are relevant for our analysis.
Apart from the $\sigma$-model terms (\ref{sigmaint}) there is a set of 
mass-like terms for the fermions and 
the scalar fields interact via the potential 
(the conventions follow \cite{N2})
\be
V=  -12 S_{AB} \bar S^{BA} + g_{i\bar j} W^{iAB}\bar W^{\bar j}_{AB}
+ 2 N_\alpha^A \bar N_A^\alpha \ ,
\end{equation}
where 
$S_{AB}, W^{iAB}, N_\alpha^A$ are scalar field dependent quantities
given by
\bea\label{Sdef}
S_{AB} &=& \frac{1}2e^{\frac{K}2} X^I  P_I^x (\sigma^x\epsilon)_{AB}
\ ,\nonumber\\
 W^{iAB} &=& i (\sigma^x\epsilon)^{AB} P_I^x g^{i\bar j} 
\bar D_{\bar j} \bar X^I\ ,\\
N_\alpha^A&=& 2 e^{\frac{K}2} U^A_{\alpha u} k^u_I \bar X^I\ .\nonumber
\end{eqnarray}
The $X^I(z)$ are holomorphic functions of the $z^i$ and their covariant
derivatives are defined as 
$D_i X^I = \partial_i X^I +(\partial_iK) X^I$ where $K$ is the 
K\"ahler potential of ${\M}_v$, i.e.\ $g_{i\bar j}=\partial_i
\bar \partial_{\bar j} K$.
The $U^A_{\alpha u}$ is the `vielbein' of the quaternionic metric
$h_{uv} = U^A_{\alpha u} U^B_{\beta v} \epsilon_{AB} C^{\alpha\beta}$
with $C^{\alpha\beta}$ being the invariant $Sp(2n_h)$ matrix.

$S_{AB}$ is also the mass matrix of the two gravitini and 
$W^{iAB}, N_\alpha^A$ are related 
to the mass matrix of the spin-$1/2$ fermions.
Furthermore, these quantities appear in the 
supersymmetry transformations of the fermions
\bea\label{susytrans}
\delta \Psi_{\mu A} &=& S_{AB} \gamma_\mu \epsilon^B +\ldots\ ,\nonumber\\
\delta \lambda^{iA} &=& W^{iAB}\epsilon_B+\ldots\ ,\\
\delta \zeta_\alpha &=& N_\alpha^A \epsilon_A+\ldots\ ,\nonumber
\end{eqnarray}
where $\gamma_\mu$ are Dirac matrices and $\epsilon^A$ are the parameters
of the two supersymmetry transformations.

\section{Spontaneous $N=2\to N=1$: the necessary conditions}

We just sketched a generic $N=2$ supersymmetric theory. It is of interest to
understand under what conditions the potential $V$ can have minima which
preserves only $N=1$ supersymmetry but not the full $N=2$.
So far there are only a few examples known where this situation 
is realized \cite{FGP,FGPT}. 

The presence of an unbroken $N=1$ supersymmetry amounts to  the requirement
that the supersymmetry transformations (\ref{susytrans}) of the fermions
evaluated
in the ground states vanish for one of the two supersymmetry
transformations, say $\epsilon_2$
\be\label{Cone}
\vev{\delta \Psi_{\mu A}} = \vev{\delta \lambda^{iA}} = 
\vev{\delta \zeta_\alpha } = 0 \ .
\end{equation}
Here the bracket $\vev{\ }$ indicates that 
the fermion transformtions have to be  evaluated
in the ground states and for $\epsilon_1 =0, \epsilon_2\neq 0$.
So the answer to the first question (i) of the introduction
amounts to determining
the properties of the couplings $S_{AB}, W^{iAB}, N_\alpha^A$
that they have to obey in order to results in a solution of (\ref{Cone}).
{}From (\ref{Sdef}) we see immediately that this is equivalent 
to geometrical conditions on the scalar manifold
${\M}$ and its (gauged) isometries.

The ground states can have various space-time properties,
they can be flat Minkowski spaces, anti-de Sitter spaces 
or an extended domain walls or $p$-brane
solutions. For simplicity we impose in the following 
the additional requirement that the ground states 
spontaneously break $N=2\to N=1$ in flat Minkowski space
leaving the more general cases to a separate publication \cite{GL}.
However, we do allow for the possibilty that there is a continuous
family of ground states which have $N=1$ supersymmetry. 
This means that there can be solutions of (\ref{Cone})
which do depend on (some of) the scalar fields. 

Solving (\ref{Cone}) directly from the definitions (\ref{Sdef})
is not straightforward. Instead we can gain a little more 
insight by further using the fact $S_{AB}$ is also
the mass matrix for the two gravitinos.
A necessary condition for the existence of $N=1$ ground states
is that the two eigenvalues $m_{\Psi^1}, m_{\Psi^2}$ of
$S_{AB}$
are non-degenerate, i.e.\ $m_{\Psi^1}\neq m_{\Psi^2}$.
In Minkowski ground states  one  also needs $m_{\Psi^2}=0$ or
in other words one of the two
gravitini has to become massive
while the second one stays massless.
Furthermore, the unbroken $N=1$ supersymmetry
implies that the massive gravitino has to be a member
of an entire $N=1$ massive spin-$3/2$ multiplet which has
the spin content $s=(3/2,1,1,1/2)$. 
This in turn requires that also two vectors,
say $A_\mu^0, A_\mu^1$
and a spin-$1/2$ fermion $\chi$ have to become massive or
equivalently there have to be two massless gauge bosons together with
two Goldstone bosons, a Goldstone fermion and a massive fermion \cite{FGPT}.
The Goldstone bosons have to be `recruited' out of a hypermultiplet
while the two gauge bosons require at least one vector multiplet.
Thus, the minimal $N=2$ spectrum which allows the possibility of a spontaneous
breaking to $N=1$ consists of the $N=2$ supergravity multiplet,
one hypermultiplet and one vector multiplet.

Since our analysis assumes Minkowskian ground states the unbroken $N=1$
supersymmtry implies that the spin-$3/2$ multiplet has to
be  degenerate in mass,
i.e.
\be\label{mdeg}
m_{\Psi^1} = m_{A^0} = m_{A^1} = m_{\chi}\equiv m\ ,\qquad
m_{\Psi^2} = 0\ .
\end{equation}

As we already said the fermionic mass matrices 
are directly determined by the 
$S_{AB}, W^{iAB}, N_\alpha^A$ defined in (\ref{Sdef}).
On the other hand the gauge bosons obtain their mass 
via a Higgs mechanism. This means that among the hypermultiplet scalars
there have to be two Goldstone bosons $\eta^1, \eta^2$ with couplings
\be\label{Gold}
D_\mu\eta^{1} = \partial_\mu\eta^{1} + e_0 A_\mu^0\ ,\qquad
D_\mu\eta^{2} = \partial_\mu\eta^{2} + e_1 A_\mu^1\ ,
\end{equation} 
where $e_0,e_1$ are constant charges and we have arbitrarily
chosen $I=0,1$ as the massive gauge bosons.
In geometrical terms this means that ${\M}_h$ has to admit
two commuting, translational ${\bf R}^2$-isometries and these 
isometries have to be gauged \cite{FGPT}.
In this case the $\sigma$-model interactions (\ref{sigmaint})
imply a set of mass terms
\be
{L} = 
- \frac12 m^2_{IJ} A_\mu^I A_\mu^J + \ldots
\end{equation}
where 
\be
\frac12 m^2_{IJ} = k_I^u h_{uv} k_J^v\ , \qquad 
k_I^u = e_0\, \delta_{I0}\delta^{u1} + e_1\, \delta_{I1}\delta^{u2}\ .
\end{equation}
The constraint (\ref{mdeg}) implies that the two Killing vectors
$k_0^u,k_1^u$ have to be orthonormal, i.e.\ satisfy
\be
k_0^u h_{uv} k_0^v = k_1^u h_{uv} k_1^v \equiv \frac12 m^2\ ,\qquad 
k_0^u h_{uv} k_1^v =0\ .
\end{equation}

We just established the fact that we need two orthonormal
Killing vector and thus via equation (\ref{Pdef})
we have two Killing prepotentials $P_0^x, P_1^x$ which generically span
a plane in  $SU(2)$. Thus, without loss of generality we can always choose 
an SU(2) basis where $P_0^3 = P_1^3 =0$. This choice fixes an $SU(2)$
gauge and leaves a $U(1)$ rotation intact. 
In this basis $S_{AB}$ is diagonal and given by
\be
S_{AB} =  \frac12 \left(\begin{array}{cc}m_{\Psi^1} &0\\ 0& m_{\Psi^2}\end{array}\right)
\ , 
\end{equation}
where
\be
m_{\Psi^1} = e^{\frac{K}2} X^I (P_I^1 - i P_I^2)\ , \qquad
m_{\Psi^2} = - e^{\frac{K}2} X^I (P_I^1 + i P_I^2)\ .
\end{equation}
Since the $P_I$ are real  we see immediately that
$m_{\Psi^2} =0$ cannot be satisfied when the $X^I$ are linearly 
independent \cite{FGP,PM1}. This implies that on ${\M}_v$
one has to choose a particular basis of gauge fields where 
the couplings (\ref{Gold}) are realized.
Leaving the detailed analysis to ref.\ \cite{GL} let us just state
that indeed such a basis exists for a large class of 
special K\"ahler manifolds ${\M}_v$ \cite{FGP,FGPT,PM1,CDFP}
and that 
\be\label{Fnot}
{\M}_v =\frac{SU(1,1)}{U(1)}\ ,\qquad \textrm{with}\qquad
X^0 = -\frac12\ ,\quad
X^1 = \frac{i}2\ ,
\end{equation}
is a representive choice \cite{FGP}. 
(Other manifolds can be found in \cite{FGPT,PM1}.)
For (\ref{Fnot}) the constraint $m_{\Psi^2} =0$ is  solved by 
\be\label{Pcon}
P_0^2 =P_1^1\ ,\qquad  P_1^2=-P_0^1\ .
\end{equation}
This is a strong constraint on the scalar manifold 
${\M}_h$ and its gauged isometries and generically defines a subspace 
of ${\M}_h$.
On this subspace one shows that (\ref{mdeg}) implies
\be
V|\equiv 0\ ,
\end{equation}
where the $|$ indicates that 
the potential is evaluated on the subspace where 
(\ref{Pcon}) holds.

Before we turn our attention to question (ii) let us summarize
the necessary conditions found so far.
The possibilty of $N=1$ ground states is equivalent to 
the existence of solutions of 
$\vev{\delta \Psi_{\mu A}} = \vev{\delta \lambda^{iA}} = 
\vev{\delta \zeta_\alpha } = 0$.
By relating it to properties of a massive $N=1$ spin-$3/2$
multiplets we were able to rephrase this as  conditions on
the scalar manifold ${\M}$. In particular on ${\M}_v$
one has to choose a basis where the $X^I$ are linearly dependent
while ${\M}_h$ has to admit two commuting, orthonormal, 
translational ${\bf R}^2$-isometries which additionally obey
(\ref{Pcon}) and (\ref{mdeg}).
It would be interesting to rephrase the constraints 
(\ref{Pcon}), (\ref{mdeg})
in a more geometrically language
and to determine the manifolds ${\M}$
which satisfy all these constraints.

\section{The low energy effective $N=1$ theory}
%
Let us now turn to question (ii) of the introduction
and focus on the properties of the 
low energy effective $N=1$ theory which
is valid well below the scale of the supersymmetry
breaking set by $m_{\Psi^1}$. The Lagrangian 
of this effective theory can be derived by `integrating out' 
the massive gravitino multiplet together with
other ($N=1$) multiplets which might have acquired 
a mass of the same order.
At the two derivative level this is achieved
by using the equation of motions of
the massive fields to first non-trivial order in $p/m_{\Psi^1}$
where $p\ll m_{\Psi^1}$ is a characteristic momentum.
For the fermions and scalars this is a straightforward procedure in that they 
are simply set to zero in the $N=2$ Lagrangian.
This in turn truncates the scalar manifold to a subspace spanned
by the left over massless states.
For the spin-1 gauge bosons the situation is slightly more complicated. 
Due to their couplings to the Goldstone bosons (\ref{Gold})
eliminating $A_\mu^{0,1}$ also elimates the two
Goldstone bosons and furthermore changes the $\sigma$-model
interactions of the remaining scalar fields.
This amounts to taking
the quotient of ${\M}_h$ with respect to the two
translational ${\bf R}^2$-isometries  \cite{HKLR}.\footnote{%
We thank G.\ Horowitz
for reminding us of this fact.}

The effective theory contains the left over massless
$N=1$ multiplets which include
a gravity multiplet, $n'_v$  vector multiplets and  
$n_c$ chiral multiplets. 
In addition, the effective theory has to be manifestly 
$N=1$ supersymmetric. This implies in particular that
the scalar manifold is K\"ahler, the gauge coupling functions $f(z)$
are holomorphic and the potential is expressed
in terms of a holomorphic superpotential $W$.
This imposes a set of 
conditions implied by the consistency of the 
integrating out procedure and 
the following (non-trivial) facts have to hold

\bi
\item[(a)]
Quaternionic-K\"ahler manifolds which admit ${\bf R}^2$-isometries 
of the type specified in section~3  have 
a quotient ${\M}_h/{\bf R}^2 $ which is K\"ahler 
(with K\"ahler potential $K_h$).\footnote{The scalar manifold for
the vector multiplets is already K\"ahler so that no new constraint arises 
here.}

\item[(b)]
The inverse gauge couplings $g$ of the gauge bosons are harmonic
$$
g^{-2} = f(z) +\bar f(\bar z)\ .
$$
\item[(c)]
The $N=1$ potential obeys
$$
V^{N=1}= e^{K_v + K_h} \left( g^{i\bar j}D_i W \bar D_{\bar j}
\bar W + g^{u\bar v}D_u W \bar D_{\bar v}
\bar W -3 |W|^2\right)\ ,
$$
where $W$ is holomorphic and 
$$
D_i W = \partial_i W +(\partial_iK_v) W\ ,
\qquad
D_u W = \partial_u W +(\partial_uK_h) W\ .
$$
($g_{u\bar v}$ denotes the K\"ahler metric on the quotient 
${\M}_h/{\bf R}^2 $.)
\ei

A generic $N=2$ theory does not satisfy  (a)--(c) 
but supersymmetry imposes these conditions on the low
energy effective $N=1$ theory.
The fact that we have chosen to consider an $N=2$ spectrum
with only one  vector multiplet
immediately implies that the low energy $N=1$ theory contains no
vector multiplets and hence (b) is trivially satisfied.
Furthermore Minkowskian ground states the $N=1$ gravitino $\Psi_\mu^2$
is exactly massless which implies $W=V^{N=1} \equiv 0$ and hence also
(c) is satisfied. Thus, for the case at hand the only non-trivial constraint 
left to check is condition~(a). 

Eliminating the two massive gauge bosons via their
equations of motions results in $\sigma$-model type couplings 
in the effective $N=1$ Lagrangian which are as in eq.\ (\ref{sigmaint})
but with $h_{uv}$  replaced by the metric $\h_{uv}$ on the quotient
given by 
\be
\h_{uv} = h_{uv} -\frac2{m^2} \big(k_{0u} k_{0v} +k_{1u} k_{1v}\big)\ ,\qquad
k_{Iu}\equiv  k^w_I h_{wu}\ .
\end{equation} 
$\h_{uv}$ satisfies
\be 
\h_{uv}k^v_I=0\ ,\qquad
\h_{uv}h^{vw}\h_{wr} = \h_{ur}\ ,
\end{equation}
where $h^{vw} h_{wu} = \delta_u^v$.
Thus  $\h_{uv}$ has two null directions and $h^{vw}$ is the inverse metric.

Among the three hyper-K\"ahler two-forms $K^3_{uv}$ plays a preferred role
in that it points in the direction (in $SU(2)$-space) normal 
to the plane spanned by $P_0^x, P_1^x$. By using a two-dimensional
$\sigma$-model one can compute the two-form 
which decends from $K^3_{wu}$ to the quotient
to be\footnote{We thank E.\ Zaslow for suggesting this procedure.}
\be
\K_{uv} = K^3_{uv} 
-\frac1k\big(k^w_0 K^3_{wu}k^w_1 K^3_{wv}-k^w_1 K^3_{wu}k^w_0 K^3_{wv}\big)\ ,
\end{equation}
where $k\equiv k^w_0 K^3_{wu}k^u_1$.
{}From (\ref{Cone}) one derives 
\be\label{crucial}
k^w_0 K^3_{wu} = -k_{1u}\ ,\qquad k^w_1 K^3_{wu} = k_{0u}\ .
\end{equation}
which in turn can be used to show\footnote{The details
of the computation will be presented in \cite{GL}. 
The proof of (\ref{juhu}) does not require the ground state to be Minkowskian.}
\be\label{juhu}
d\K = 0 \ , \qquad \J^2 = - {\bf 1}\ ,
\end{equation}
where $\K_{uv} = \h_{uw} \J^w_v$.
This proves that the quotient is indeed a K\"ahler manifold 
with K\"ahler form $\K$ and complex structure $\J$.\footnote{It is 
tempting to conjecture that this K\"ahler manifold is somehow
related to the twistor space \cite{HKLR,WRV}. 
We thank P.\ Aspinwall, S.\ Vandoren and B.\ de Wit for discussions 
related to this conjecture.}
Hence the consistency condition (a) is satisfied.

Let us close by summarizing the properties of the K\"ahler manifold 
just constructed. We started from a quaternionic 
manifold ${\M}_h$ which admits two orthonormal 
Killing vectors of an ${\bf R}^2$-isometry.
We showed that if in addition (\ref{crucial}) holds the quotient manifold 
${\M}_h/{\bf R}^2$ is K\"ahler.
It would be interesting to determine the quaternionic geometries
which do satisfy (\ref{crucial}) and thus (\ref{juhu}).

\vskip 1cm

{\bf Acknowledgments}

This work is supported in part by
the German Science Foundation (DFG),
the German--Israeli
Foundation for Scientific Research (GIF),
the European RTN Program HPRN-CT-2000-00148 and 
the German Academic Exchange Service (DAAD).

It is a great pleasure to thank
P.~Aspinwall, B.~de Wit, S.~Ferrara, J.~Gheerardyn, B.~Gunara,
G.~Horowitz, V.~Kaplunovsky, P.~Mayr, G.~Moore, F.~Roose, A.~Strominger, 
S.~Theisen, C.~Vafa, S.~Vandoren, A.~Van Proeyen, E.~Zaslow
for helpful discussions and D.V.~Alekseevsky, V.~Cort\'es, C.~Devchand, 
A.~Van~Proeyen for organizing a fruitful and stimulating
workshop.

\vskip 1cm



\begin{thebibliography}{99}
%
\bibitem{HP}
J.~Hughes and J.~Polchinski,
``Partially Broken Global Supersymmetry And The Superstring'',
Nucl.\ Phys.\  {\bf B278} (1986) 147;\\
J.~Hughes, J.~Liu and J.~Polchinski,
``Supermembranes'',
Phys.\ Lett.\  {\bf B180} (1986) 370.
%
\bibitem{BG}
J.~Bagger and A.~Galperin,
``Matter couplings in partially broken extended supersymmetry'',
Phys.\ Lett.\  {\bf B336} (1994) 25, hep-th/9406217;\\
``A new Goldstone multiplet for partially broken supersymmetry'',
Phys.\ Rev.\  {\bf D55} (1997) 1091, hep-th/9608177;\\
``Linear and nonlinear supersymmetries'',
hep-th/9810109.
%
\bibitem{APT}
I.~Antoniadis, H.~Partouche and T.~R.~Taylor,
``Spontaneous Breaking of N=2 Global Supersymmetry'',
Phys.\ Lett.\  {\bf B372} (1996) 83, hep-th/9512006.
%
\bibitem{FGP}
S.~Ferrara, L.~Girardello and M.~Porrati,
``Minimal Higgs Branch for the Breaking of Half of the Supersymmetries in N=2 Supergravity'',
Phys.\ Lett.\  {\bf B366} (1996) 155, 
hep-th/9510074;\\
``Spontaneous Breaking of N=2 to N=1 in Rigid and Local Supersymmetric Theories'',
Phys.\ Lett.\  {\bf B376} (1996) 275, hep-th/9512180.
%
\bibitem{FGPT}
P.~Fre, L.~Girardello, I.~Pesando and M.~Trigiante,
``Spontaneous N = 2 $\to$ N = 1 local supersymmetry breaking with surviving  compact gauge groups'',
Nucl.\ Phys.\  {\bf B493} (1997) 231, hep-th/9607032.
%
\bibitem{Bac}
C.~Bachas, ``A way to break supersymmetry'',
{\tt hep-th/9503030}.
%
\bibitem{JM} J.~Michelson,
``Compactifications of Type IIB Strings to Four 
Dimensions with  
non-trivial Classical Potential,''
Nucl.\ Phys.\  {\bf B495} (1997) 127,
hep-th/9610151.
%
\bibitem{KK}
E.~Kiritsis and C.~Kounnas,
``Perturbative and non-perturbative partial supersymmetry breaking:  N = 4 $\to$ N = 2 $\to$ N = 1'',
Nucl.\ Phys.\  {\bf B503} (1997) 117, hep-th/9703059.
%
\bibitem{TV}T.\ Taylor and C.~Vafa,
``RR Flux on Calabi-Yau and Partial Supersymmetry Breaking'', 
Phys.\ Lett.\  {\bf B474} (2000) 130,
hep-th/9912152.
%
\bibitem{PM1}
P.~Mayr,
``On supersymmetry breaking in string theory and its realization in brane  worlds'',
Nucl.\ Phys.\  {\bf B593} (2001) 99, hep-th/0003198.
%
\bibitem{CKLT}
G.~Curio, A.~Klemm, D.~L\"ust and S.~Theisen,
``On the vacuum structure of type II string compactifications on  Calabi-Yau
spaces with H-fluxes'', Nucl.\ Phys.\  {\bf B609} (2001) 3,
hep-th/0012213.
%
\bibitem{ADF}
L.~Andrianopoli, R.~D'Auria and S.~Ferrara,
``Supersymmetry reduction of N-extended supergravities in four  dimensions'',
hep-th/0110277;\\
``Consistent reduction of N = 2 $\to$ N = 1 four dimensional supergravity  coupled to matter'',
hep-th/0112192.
%
\bibitem{ADFL}
L.~Andrianopoli, R.~D'Auria, S.~Ferrara and M.~A.~Lledo,
``Super Higgs effect in extended supergravity'', hep-th/0202116.
%
\bibitem{GL}
B.E.\ Gunara and J.\ Louis, in preparation.



\bibitem{N2}
For a review of gauged $N=2$ supergravity
see, for example,\\
L.~Andrianopoli, M.~Bertolini, A.~Ceresole, R.~D'Auria, S.~Ferrara, P.~Fre and T.~Magri,
``N = 2 supergravity and N = 2 super Yang-Mills theory on general scalar  manifolds: Symplectic covariance, gaugings and the momentum map'',
J.\ Geom.\ Phys.\  {\bf 23} (1997) 111, hep-th/9605032, and references
therein.
%
\bibitem{DP} 
B.~de~Wit and A.~Van~Proeyen, ``Potentials and symmetries of general gauged
  {N=2} supergravity - {Y}ang-{M}ills models'', Nucl.\ Phys.\ {\bf B245}
  (1984) 89.
%
\bibitem{CRTP} 
For a review see, for example, 
B.~Craps, F.~Roose, W.~Troost and A.~Van Proeyen,
``What is special Kaehler geometry?'',
Nucl.\ Phys.\ B {\bf 503} (1997) 565, hep-th/9703082.
%
\bibitem{BW}
J.~Bagger and E.~Witten, ``Matter couplings in {N=2} supergravity'', 
Nucl.\  Phys.\ {\bf B222} (1983) 1;\\
B.~de~Wit, P.~G. Lauwers, and A.~Van~Proeyen, ``Lagrangians of {N=2}
  supergravity - matter systems'', Nucl.\ Phys.\ {\bf B255} (1985) 569.
%
\bibitem{DFF}
K.~Galicki,
``A Generalization Of The Momentum Mapping Construction For Quaternionic Kahler Manifolds'',
Commun.\ Math.\ Phys.\  {\bf 108} (1987) 117;\\
R.~D'Auria, S.~Ferrara, and P.~Fre, ``Special and quaternionic isometries:
  General couplings in N=2 supergravity and the scalar potential'', 
Nucl.\  Phys.\ {\bf B359} (1991) 705.


\bibitem{CDFP}
A.~Ceresole, R.~D'Auria, S.~Ferrara and A.~Van Proeyen,
``Duality transformations in supersymmetric Yang-Mills theories coupled to supergravity'',
Nucl.\ Phys.\ B {\bf 444} (1995) 92, hep-th/9502072.
%
\bibitem{HKLR}
N.~J.~Hitchin, A.~Karlhede, U.~Lindstrom and M.~Rocek,
``Hyperkahler Metrics And Supersymmetry'',
Commun.\ Math.\ Phys.\  {\bf 108} (1987) 535.
%

\bibitem{WRV}
B.~d.~Wit, M.~Rocek and S.~Vandoren,
``Gauging isometries on hyper-K\"ahler cones 
and quaternion-K\"ahler  manifolds'',
Phys.\ Lett.\ B {\bf 511} (2001) 302, hep-th/0104215;\\
``Hypermultiplets, hyperk\"ahler cones and quaternion-K\"ahler geometry'',
JHEP {\bf 0102} (2001) 039, hep-th/0101161.


\end{thebibliography}
\end{document}